\documentclass{PoS}
\usepackage{bm}

\def\({\left(}
\def\[{\left[}
\def\){\right)}
\def\]{\right]}

\title{Layout of Wilson lines and light-cone peculiarities of transverse-momentum dependent PDFs}

\ShortTitle{Layout of Wilson lines}

\author{\speaker{I.O. Cherednikov}\thanks{On leave of absence from the Bogoliubov Laboratory of Theoretical Physics, JINR, 141980 Dubna, Russia}\\
        Departement Fysica, Universiteit Antwerpen, B-2020 Antwerpen, Belgium\\
        E-mail: \email{igor.cherednikov@ua.ac.be}}

\abstract{We consider the problems of gauge invariance, path-dependence and treatment of overlapping UV/rapidity divergences peculiar to the transverse-momentum dependent parton distribution functions (TMDs). For different formulations of the TMDs available in the literature, we check the consistency of the TMD matrix elements with the collinear parton distribution functions possessing the well-known operator structure. Comparative on- and off-light-cone layout of the Wilson lines which secure the gauge-invariance of the TMDs is presented and briefly discussed.}

\FullConference{Sixth International Conference on Quarks and Nuclear Physics,\\
		April 16-20, 2012\\
		Ecole Polytechnique, Palaiseau, Paris}

\begin{document}

Collinear (transverse-momentum-integrated) parton densities can be introduced as the gau\-ge\--in\-de\-pen\-dent hadronic matrix elements \cite{DIS_p} (light-cone components of four-vectors are defined as $p^\pm = (p^0 \pm p^z)/{\sqrt 2}$)
\begin{equation}
  f_a(x, \mu^2)
=
  \frac{1}{2}
  \int \frac{d\xi^- }{2\pi }
  \ {\rm e}^{-ik^{+}\xi^{-} }
  \left\langle
              p\  |\bar \psi_a (\xi^-, \mbox{\boldmath$0_\perp$})[\xi^-, 0^-]_n
              \gamma^+
   \psi_a (0^-,\mbox{\boldmath$0_\perp$}) | \ p
   \right\rangle \ .
   \label{eq:iPDF}
\end{equation}
Generic Wilson (straight) line $[\xi^-, 0^-]_w$ is a path-ordered exponential evaluated along the direction of a fixed four-vector $w$:
$[\infty; \xi]_{w}
\equiv {}
  {\cal P} \exp \left[
                      - i g \int_0^\infty d\tau \ w_{\mu} \
                      A_{a}^{\mu}t^{a} (\xi + w \tau)
                \right] \
                $\,.
For fully inclusive processes, the relevant integration path is completely defined by a single light-like vector $n$.
The matrix element (\ref{eq:iPDF}) is, therefore, gauge invariant, but path-($n-$)dependent. This path-dependence, however, does not give rise to any problems in the case of the collinear PDFs where the situation is effectively one-dimensional \cite{CFP80}.

Application of the QCD factorization approach to the semi-inclusive processes, such as semi-inclusive deep inelastic scattering (SIDIS), the Drell-Yan (DY) process etc., requires more detailed knowledge of the nucleon structure: one must be able, in particular, to take into account intrinsic transverse momenta of the partons and their possible correlations with the spin of the parent nucleon. To this end, the transverse-momentum dependent (TMD) distribution and fragmentation functions have been proposed (for a review, see, e.g., \cite{TMD_INT, TMD_LHC} and refs. therein). To our discussion, we adopt the following definition of the trial unsubtracted``quark in a quark'' TMD which is a straightforward generalization of the collinear matrix element (\ref{eq:iPDF}):
\begin{eqnarray}
&& {\cal F}_{\rm unsub.} \left(x, {\bm k}_\perp; \mu \right)
=
  \frac{1}{2}
  \int \frac{d\xi^- d^2 {\xi}_\perp}{2\pi (2\pi)^2} \
  {\rm e}^{-ik \cdot \xi}
  \left\langle
              p \ |\bar \psi_a (\xi^-,  \bm{\xi}_\perp)
              [\xi^-,  \bm{\xi}_\perp;
   \infty^-,  \bm{\xi}_\perp]_{n}^\dagger  \right.  \nonumber \\
   && \left.
\times
   [\infty^-,  {\xi}_\perp;
   \infty^-,  {\infty}_\perp]_{\bm l}^\dagger
   \gamma^+[\infty^-,  {\infty}_\perp;
   \infty^-, \bm{0}_\perp]_{\bm l}
   [\infty^-, \bm{0}_\perp; 0^-,\bm{0}_\perp]_{n}
   \psi_a (0^-,\bm{0}_\perp) | \ p
   \right\rangle\Big|_{\xi^+=0} \ .
\label{eq:general}
\end{eqnarray}
We distinguish here between longitudinal $n$- and transversal $\bm l$- Wilson lines \cite{TMD_LC_trans}.
Formally, performing integration of the Eq. (\ref{eq:general}) over $\bm k_\perp$ one expects to obtain the standard collinear PDF, Eq. (\ref{eq:iPDF})
\begin{equation}
  \int\! d^2 \bm k_\perp \ {\cal F}_{\rm unsub.} (x, \bm k_\perp)
  =
 \frac{1}{2}
  \int \frac{d\xi^- }{2\pi } \
  {\rm e}^{-ik^{+}\xi^{-} } \
  \left\langle
              p\  |\bar \psi_a (\xi^-, \bm 0_\perp)[\xi^-, 0^-]_n
              \gamma^+
   \psi_a (0^-,\bm 0_\perp) | \ p
   \right\rangle \ = f_a(x) \ .
   \label{eq:u_to_i}
\end{equation}
We focus first on the ``extended tree-level'' picture: we do not take into account the quantum corrections due to perturbative gluon exchanges, but keep on with the ``classical'' Wilson lines. Staying within this approximation, we can still learn something about the geometrical structure of the Wilson lines in the TMDs. The dependence on the integration path is a crucial point here.

While the fermionic structure of the quark TMDs is fixed (given the parton number interpretation, it must be a bilocal product of two quark field operators defined in two different space-time points separated by a non-light-like interval), there are several different approaches to the design of the Wilson lines which resum collinear gluons and provide the gauge independence of the TMD matrix elements. Moreover, another set of the Wilson lines arises in the so-called soft factors, which enable one to get rid of the extra rapidity divergences. We present and discuss the geometry of the gauge links in different formulations of the quark TMD in Figs. 1-3.

\begin{figure}[h]
\begin{center}
\includegraphics[width=0.4\textwidth,height=0.60\textheight,angle=90]{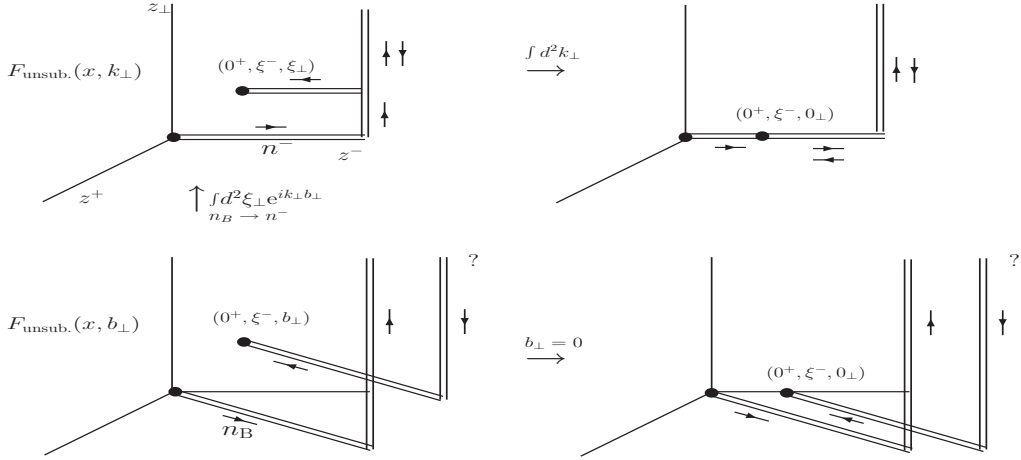}
\caption{\label{fig:1} Layout of the integration paths in the trial quark TMD with the light-like (upper left panel) and off-the-light-cone (lower left panel) longitudinal Wilson lines; the symbolic reduction to the collinear PDFs (right panels). In the case of the fully light-like longitudinal Wilson lines, the transverse gauge links at the light-like minus-infinity disappear after the integration over the intrinsic transverse momentum. The longitudinal Wilson lines produce, in the end, the straight connector $[\xi^-, 0^-]$ lying on the minus-ray (for detailed discussion, see \cite{Ste83}). The off-the-light-cone case: the transverse Wilson lines at the minus-infinity apparently do not cancel. Additionally, the collinear picture restored from the off-the-light-cone TMD consists of the two semi-infinite off-the-light-cone Wilson lines which are not equal to the straight connector $[\xi^-, 0^-]$, as it is needed for the DGLAP evolution. The renormalization group properties of these two TMDs are also different. The interrogation marks mean that the consistent treatment of the transverse Wilson lines at the minus-infinity in the TMDs with off-the-light-cone longitudinal Wilson lines is still lacking. In contrast, the transverse Wilson lines are quite natural within the pure ``light-cone'' approaches.}
\end{center}
\end{figure}

\begin{figure}[h]
\begin{center}
\includegraphics[width=0.4\textwidth,height=0.60\textheight,angle=90]{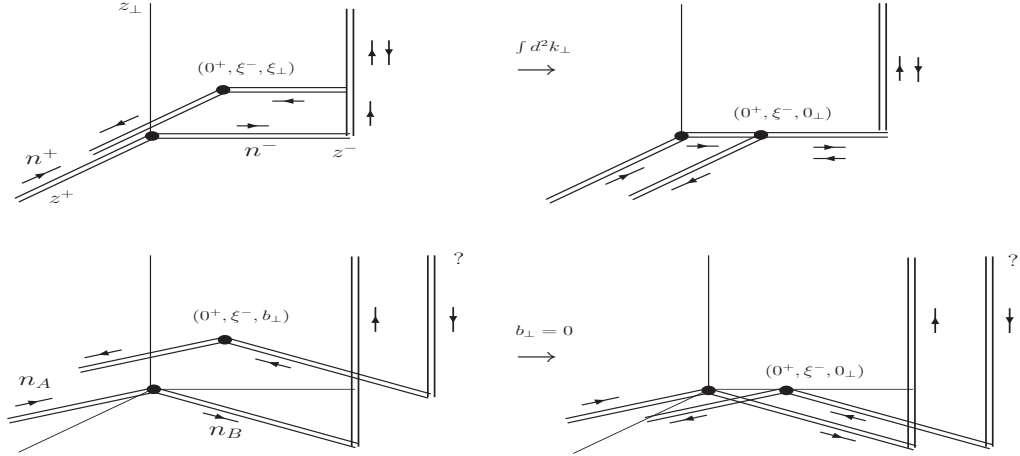}
\caption{\label{fig:2}:Layout of the Wilson lines entering the ``bare'' soft factors (left column) and  the symbolic reduction to the integrated PDFs (right column). The upper panel shows the soft factor in the momentum space proposed in \cite{CS_all}. The lower panel presents the ``shifted'' off-the-light-cone contour in the impact parameter space and the reduction to the collinear PDF at $\bm b_\perp \to 0$.}
\end{center}
\end{figure}

\begin{figure}[h]
\begin{center}
\includegraphics[width=0.4\textwidth,height=0.6\textheight,angle=90]{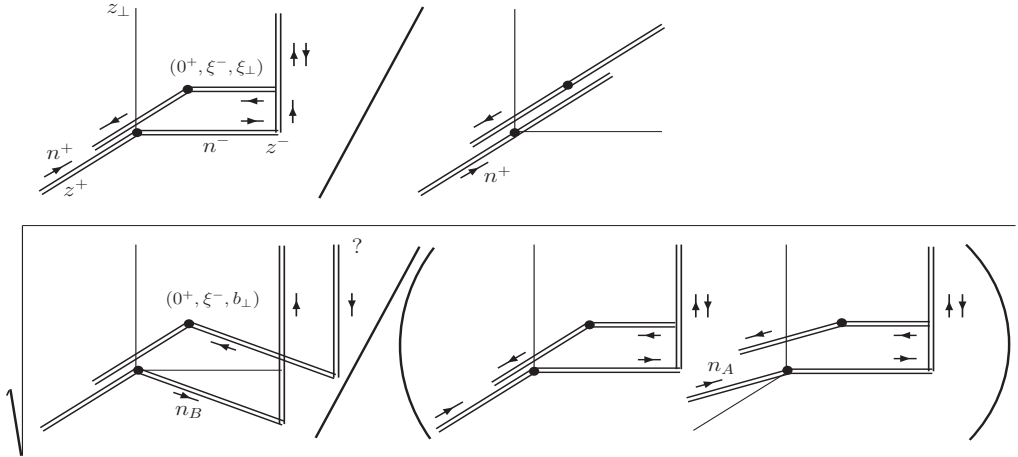}
\caption{\label{fig:3}Comparative structure of the Wilson lines in the ``full'' soft factors. The upper panel shows the soft factor of the ``pure light-cone'' TMD \cite{CS_all, ChSt_recent}. The lower panel shows the longitudinal gauge links shifted from the light-cone, which are used in the Collins factorization approach \cite{New_TMD_Col, New_TMD_PHENO}.
}
\end{center}
\end{figure}

\begin{figure}[h]
\begin{center}
\includegraphics[width=0.35\textwidth,height=0.35\textheight,angle=90]{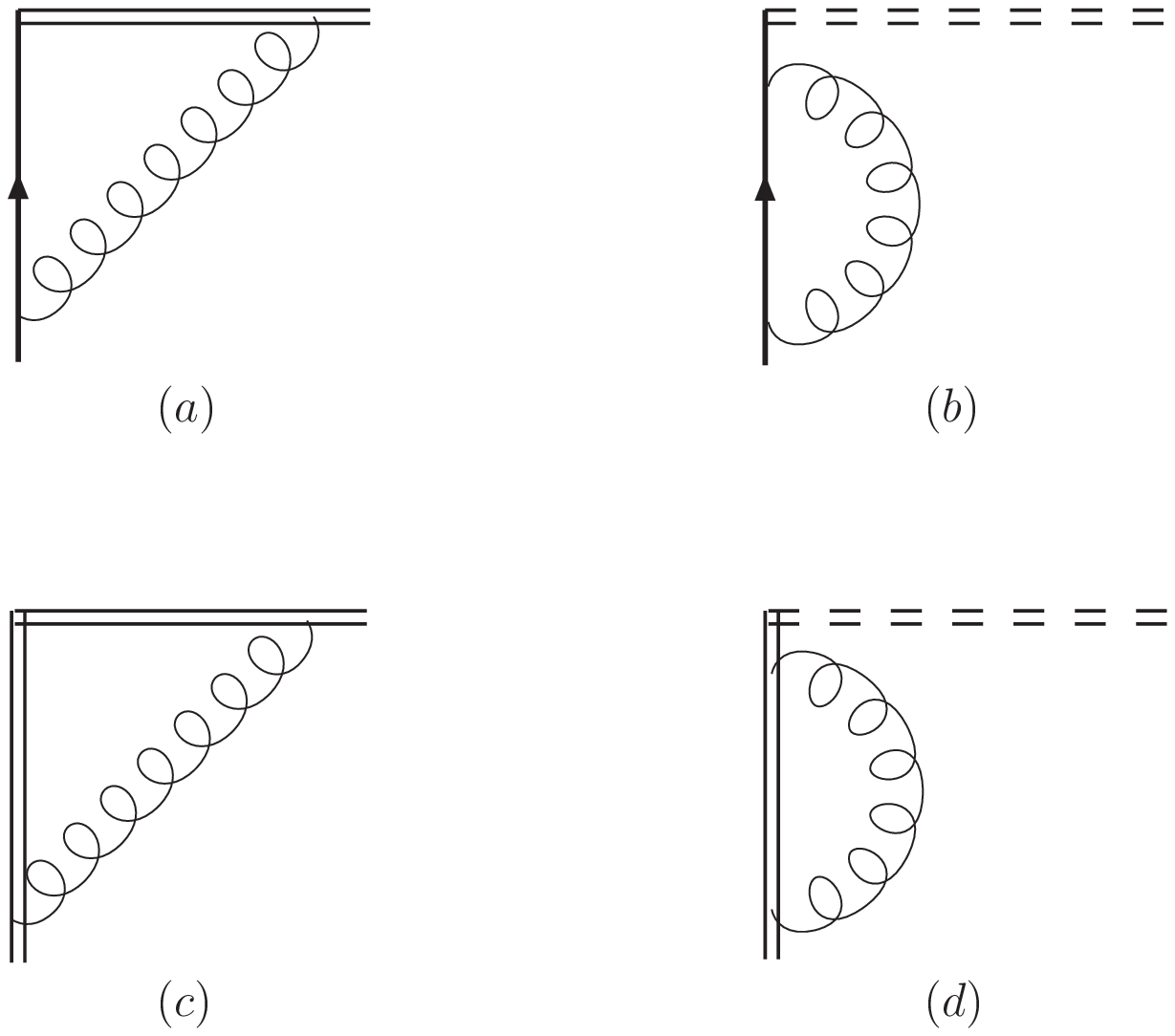}
\caption{\label{fig:4}The virtual one-loop Feynman graphs which produce overlapping UV/rapidity singularities: $(a)$---vertex-type fermion-Wilson line interaction in covariant gauge; $(b)$---self-energy graph which produces the rapidity divergences in the light-cone gauge with the advanced pole prescription (there are no transverse gauge links in this case); $(c,d)$ are the counter-parts of $(a,b)$ from the soft factor.}
\end{center}
\end{figure}

Consider now the overlapping singularities emerging in the one-loop corrections to the trial TMD (\ref{eq:general}). The corresponding Feynman graphs are presented in Fig. 4. In the covariant gauge, the UV-singular term with potential rapidity divergency reads
\begin{equation}
  { \Sigma^{(a)} }
=
- \frac{\alpha_s}{\pi} C_{\rm F}
 \Gamma (\epsilon) \left[ 4\pi \frac{\mu^2}{-p^2} \right]^{\epsilon} \
 { \int_0^1\! dx \ \frac{x^{1-\epsilon}}{(1-x)^{1+\epsilon}} } \ .
\end{equation}
Its light-cone gauge counter-part is
\begin{equation}
  {\Sigma^{(b)}}
=
  - \frac{\alpha_s}{\pi} C_{\rm F} \  \Gamma(\epsilon)\
  \left[ 4 \pi \frac{\mu^2}{-p^2} \right]^\epsilon\
  \ { \int_0^1\!
  dx \frac{(1-x)^{1-\epsilon}}{x^{1+\epsilon} } } \ .
\end{equation}
In both cases, besides the dimensional regularization, an additional cutoff is needed to control the rapidity divergences peculiar to the non-local matrix elements with light-like Wilson lines (or to those ones taken in the light-cone axial gauge).
On the other hand, the soft factor (lower panel) gives the following one-loop contributions
\begin{equation}
  \Sigma^{(c)} = \Sigma^{(d)}
  = - \frac{\alpha_s}{\pi} C_{\rm F} \  \Gamma(\epsilon)\
  \left[ 4 \pi \frac{\mu^2}{\lambda^2} \right]^\epsilon\
  { \int_0^1\!
  dx \frac{1}{x (1-x) } }  \ .
\end{equation}
Adopting the $\bar \eta$-regularization of the gluon propagator in the light-cone gauge \cite{CS_all}, or, equivalently, the similar regularization of the eikonal propagator in the covariant gauge \cite{SCET_TMD}, we obtain the UV-singular terms
\begin{equation}
  \Sigma_{\rm UV}^{(a,b)} (\epsilon, \bar \eta)
  =
  \frac{\alpha_s}{\pi} C_{\rm F}\ \frac{1}{\epsilon} \ \ln \bar \eta \  \ , \ \
  \Sigma_{\rm UV}^{(c,d)} (\epsilon, \bar \eta, \bar \eta_+)
  =
  - \frac{\alpha_s}{\pi} C_{\rm F}\ \frac{1}{\epsilon} \ \ln {\bar \eta}{\bar \eta_+} \ .
\end{equation}
where the dimensionless regulator reads $\bar \eta = \eta/p^+$, and $\bar \eta_+$ is an additional regulator of the plus-light-cone divergences in the soft factor \cite{CS_all, SCET_TMD}. Since we are interested in the TMD evolution with respect to the rapidity scale $\theta = \ln \eta/p^+$, it is convenient to take the ``rapidity derivative'' to get rid of the light-cone-driven peculiarity \cite{K_LC}, see also discussions in \cite{TMD_LC_sing}:
\begin{equation}
  \frac{d \Sigma^{(a,b)}_{\rm UV}}{d \theta} =
  - \frac{d \Sigma^{(c,d)}_{\rm UV}}{d \theta}
  =
  \frac{\alpha_s}{\pi} C_{\rm F} \ \frac{1}{\epsilon} \ .
\end{equation}
The residues of the $\Sigma$'s at a simple pole $1/\epsilon$ are given, therefore, by the plus and minus cusp anomalous dimension for a light-cone angle \cite{KR87, K_LC}:
$
  \gamma_{\rm cusp} = \frac{\alpha_s}{\pi} C_{\rm F}
$.
As for the generic structure of the overlapping divergences in the TMDs in the next-to-leading orders, we conjecture that
the contribution of the overlapping singularities to the renormalized TMD is defined by (in general, a finite number of) the cusp anomalous dimensions which are known in the theory of Wilson lines/loops, and the cancellation of those singularities is achieved by the proper subtraction of the {soft factors with obstructions (angles)}. We demonstrated the validity of this conjecture in the leading $O(\alpha_s)$-order in our previous works \cite{CS_all, ChSt_recent}. The symbolic factorization formula which is assumed in this approach reads
\begin{equation}
    W^{\mu\nu}
 =
 |H(Q,\mu)^2|^{\mu\nu} \cdot \frac{{\cal F}_{\rm unsub.}^{[{\rm A}_{\rm n}]} (x, \bm k_\perp; \mu, \theta)}{S_F(n^+,n^-; \theta) L_F^{-1}(n^+)} \otimes \frac{{\cal D}_{\rm unsub.}^{[{\rm A}_{\rm n}]} (z, z \bm k'_\perp; \mu, \theta)} {S_D(n^+,n^-; \theta) L_D^{-1}(n^-)} + ... \ .
 \label{eq:LC_factor}
\end{equation}
The layout of the Wilson lines is consistent with that of the integrated PDFs \cite{K_LC, Li_97} and allows the parton number interpretation.

On the other hand, a new design of the generic TMD has been proposed by Collins \cite{New_TMD_Col}. In his approach, the TMDs are defined in the impact parameter space and contain the soft factors with
the Wilson lines evaluated along the different light-like $n^\pm$ and the ``shifted'' $n_{A,B}$ vectors (see also \cite{JMY})
\begin{equation}
  {\cal F}^{\rm [Col.]} (x, \bm b_\perp; \mu, \zeta_F)
  =
  {\cal F}_{\rm unsub.}^{[{\rm A}_{\rm n}]} (x, \bm b_\perp; \mu) \cdot \sqrt{\frac{S(n^+,n_B)}{S(n^+,n^-)S(n_A,n^-)}} \ .
\label{eq:TMD_Col}
\end{equation}
The structure of these Wilson lines is shown and discussed in Fig. 3.

Another interesting approach to the TMD factorization and treatment of the rapidity divergences has been recently proposed in Refs. \cite{SCET_TMD}. Let us point out that there is still no visible connection between different formulations of TMDs. In particular, the operator definitions of the TMDs may correspond to different (although, somehow related) objects with different renormalization and evolution properties. In each case, however, the issue of path-dependence being hidden in the collinear PDFs appears to be crucial to understanding of the operator structure of the gauge-invariant non-local matrix elements.

\paragraph{Acknowledgements}
This work is based on the results obtained in collaboration with N.G. Stefanis, to whom I am grateful also for numerous fruitful discussions on the subject. I thank the Organizers of the conference QNP 2012 for the hospitality and warm atmosphere during the conference.

\end{document}